\documentclass[twocolumn,showpacs,preprintnumbers,amsmath,amssymb]{revtex4}

\usepackage{graphicx}

\begin{document}

\title{Thermal denaturation of an helicoidal DNA model}

\author{Maria Barbi}
\email{barbi@lptl.jussieu.fr}
\affiliation{Laboratoire de Physique Th\'eorique des Liquides, \\
Universit\'e "P. et M. Curie'', 4 Place Jussieu,
75252 Paris Cedex 05, France}

\author{Stefano Lepri}
\email{stefano.lepri@unifi.it}
\affiliation{Istituto Nazionale per la Fisica della Materia, 
UdR Firenze, via G. Sansone 1 50019 Sesto Fiorentino, Italy}
\author{Michel Peyrard}
\email{Michel.Peyrard@ens-lyon.fr}
\affiliation{Laboratoire de Physique, \'Ecole Normale Sup\'erieure de Lyon, \\
46 All\'ee d'Italie, 69364 Lyon Cedex 07, France} 
\author{Nikos Theodorakopoulos}
\email{nth@eie.gr}
\affiliation{Theoretical and Physical Chemistry Institute,\\
National Hellenic Research Foundation\\
Vasileos Constantinou 48, Athens 116 35, Greece.}

\begin{abstract}

We study the static and dynamical properties of DNA in the vicinity of its
melting transition, i.e.\ the separation of the two strands upon heating. The
investigation is based on a simple mechanical model which includes the
helicoidal geometry of the molecule and allows an exact numerical evaluation
of its thermodynamical properties. Dynamical simulations of long-enough
molecular segments allow the study of the structure factors and of the
properties of the denaturated regions. Simulations of finite chains display
the hallmarks of a first order transition for sufficiently long-ranged
stacking forces although a study of the model's ``universality class''
strongly suggests the presence of an ``underlying'' continuous transition.

\end{abstract}




\maketitle
 
\section{Introduction}

The study of simple dynamical models describing various features of
DNA dynamics has interested many authors for almost 20 years
\cite{Yak}. This vivid interest arises of course from the biological
relevance of DNA but also from its physical properties which can now
be probed through single-molecule micromanipulation experiments
like stress-induced transitions \cite{Str99} or strand separation
\cite{Boc98}.  This series of studies has clearly
pointed out that DNA must be considered as a dynamical object, whose
(nonlinear) distortions could play a major role in its functions.  

One feature of DNA that attracted a lot of attention from
physicists is its thermal denaturation, i.e.\ the transition
from the native double-helix B-DNA to its melted form where the two
strands spontaneously separate upon heating \cite{War85}, because it
provides an example of a {\em one-dimensional phase transition.}
Experiments show that this transition is very sharp, which suggested
that it could be first order and this led to numerous investigations,
first to justify the existence of a transition in a one-dimensional
system and second to determine its order
\cite{ZIMM,AZBEL,AZBELL79,The00,Dau02,Cau00,Muk00,Car02}.
In spite of these efforts the nature of the transition is still
unclear and moreover, as discussed below, the determination of its
``true'' order from experiments may turn out to be virtually
impossible. 
 
In the theoretical approaches, the level of complexity is reduced to the
minimum by taking into account only the (classical) motion of large
subunits rather than the full (quantum) many-body dynamics of all the
atoms. Clearly, an appropriate choice of the relevant degrees of
freedom, depending on the specific problem at hand, is crucial. Models
based on the theory of polymers use self-avoiding walks to describe the
two strands \cite{Cau00,Muk00,Car02}. They can be very successful in
studying the properties of the melting transition at the largest scale
but, as they do not describe DNA at the level of the base pairs, they
cannot be used to investigate properties that depend on the sequence, or
probe DNA at a microscopic scale such as some recent single-molecule
experiments.  One of the simplest models that investigates DNA at the
scale of a base pair is the Peyrard-Bishop (PB) model
\cite{Pey89,Dau93,Dau95}.  
The complex double-stranded molecule is
described by postulating some simple effective interaction among the
bases within a pair and along the strands. The model has been
successfully applied to analyze experiments on the melting of short
DNA chains \cite{Cam98}. Furthermore, it allows to easily include the
effect of heterogeneities \cite{Hwa97} yielding a sharp staircase
structure of the melting curve (number of open base pairs as a
function of the temperature $T$) \cite{War85}. Beyond its original
motivation to explain the denaturation transition, the PB model has an
intrinsic theoretical interest  as one of the simplest
one-dimensional systems displaying a genuine phase transition
\cite{The00,Dau02}.

The PB model has however a serious shortcoming because it does not take into
account the helicoidal structure of DNA. In the following, we consider an
extended version of the PB model that has been proposed to avoid this weakness
\cite{Bar99a,Bar99b}.  The introduction of DNA geometry  induces an important
coupling between base pair opening and helical twist, largely substantiated for
real DNA \cite{Bar99a}.  A modified version of this helicoidal model has also
been successfully applied to describe the denaturation  of the chain induced
either thermally or mechanically by applying an external torque to the chain ends
\cite{Coc99b}. Furthermore, its nonlinear excitations have  been studied:  small
amplitude breather-like solutions have been analytically determined \cite{Coc99a}
as well as large localized bubbles \cite{Cam01}. Both types of excitations are of
interest as they may be thermally excited as precursors for the DNA strand
separation.  A first effect of the helicoidal structure, namely to bring close to
each other bases which are not consecutive in the sequence, was treated in
\cite{Gaeta}, by introducing an interaction between these bases. This first
approach however did not take into account the important geometrical effect that
we want to examine here, the coupling between opening and twist.

The aim of this work is to give further insight on the melting transition of
the helicoidal model, both from the statistical and from the dynamical point
of view. After having recalled the model and its state variables (Sections
\ref{sec:model} and \ref{sec:qualitative}),  the first part (Sections
\ref{sec:transferint} and \ref{sec:md}) is devoted to
its exact thermodynamics and to a simulation study of its statistical
properties. We find that the melting transition is extremely sharp,
bearing essentially all of the hallmarks of a first-order 
transition, at all temperature sampling steps studied (down to 0.01~K).
However, a study of the model's ``universality class'', 
using finite-size scaling techniques, allowing some variation of the 
relevant physical parameters, and drawing from analogies with
a Schr\"odinger-like equation (Appendix A), strongly suggests
the presence of an ``underlying'' continuous transition 
of the Kosterlitz-Thouless type in the absence of nonlinear stacking.
This should be contrasted with the exact second-order transition
obtained in the zero-stacking limit of the PB model \cite{Dau02}.

In the second part (Section \ref{sec:dynamics}) we investigate
dynamical structure factors that are of particular relevance to both
neutron \cite{Gri97} and Raman \cite{Ur91} scattering experiments.
The underlying idea is that, upon approaching the transition, the
molecule should display precursor effects in the form of a "mode
softening", i.e.\ a slowing-down of the dynamics with appearance of a
low-frequency component in the spectrum. Slowing effects have been, to
some extent, observed in the structure factors of the PB model
\cite{Dau95} thus encouraging this type of investigation.

\section{The helicoidal model}
\label{sec:model}

\begin{figure}[uh]
\includegraphics[clip,width=80mm]{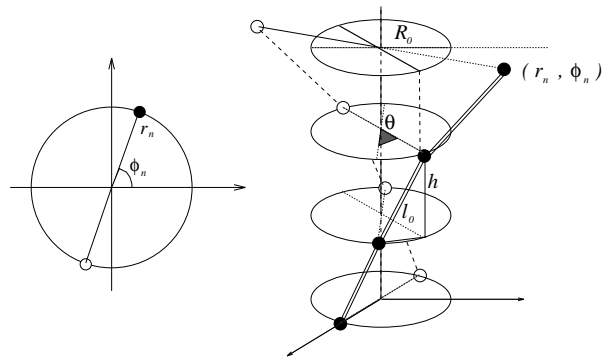}
\caption{Schematic representation of the fixed-planes DNA helicoidal model.}
\label{fig:model}
\end{figure}

Two versions of the model have been proposed in earlier studies.  The first
one \cite{Bar99a} describes an elastic backbone and fixed base-pair planes
while the second \cite{Coc99b} considers a rigid backbone and moving
base-pair planes.  The two models display a very similar behavior with
respect to denaturation as the potentials associated to the base-base
interactions in a pair and along the strands are the same  in both cases,
and because both introduce a coupling between opening and twist that
results from the helicoidal geometry. In the following, we will consider
the fixed-planes model, sketched in Fig.~\ref{fig:model}.

The helical structure of DNA is introduced essentially
by the competition between a stacking interaction that tends to keep
the base-pairs close to each other (given by the fixed
distance $h$ between the base planes) 
and the length $\ell_0 > h$ of the backbone segment
(described as an elastic rod of rest length $\ell_0$)
that connects the attachment points of the bases along each strand.
The ratio $\ell_0/h$ fixes the  
strand slant and therefore the resulting helicity of the
structure. This helicity is accounted for by the angle of rotation of
a base-pair with respect to the previous one, namely the {\em
equilibrium twist angle} $\theta$, equal to $2\pi/10.4$ in B-DNA at
room temperature.

In the model, bases are described as point-like particles of equal masses $m$
joined by elastic rods along each strand. Bases lying on the same plane are
coupled by hydrogen bonds leading to an attractive force that tends to maintain
their equilibrium distance equal to the DNA diameter $2R_0$.  We assume that
the two bases in each pair move symmetrically.  To describe base-pair opening
and helical torsion under those constraints, it suffices to introduce two
degrees of freedom per base-pair: these are $r_n$ and $\phi_n$, i.e.\ the
radial and angular  positions and of the $n-$th base. The total number of
degrees of freedom is thus $2N$, $N$ being the total number of base pairs. The
restriction imposed by the assumption of a symmetric motion preserves the
essential feature of the helical structure, the coupling between torsion and
opening, while it keeps the model sufficiently simple to allow an {\em exact}
treatment of its thermodynamics.

We consider the Lagrangian \cite{note}
\begin{eqnarray}
\label{lagrangian}
\cal{L} &=& m \sum_n \Big(\dot{r}_n^2 + r_n^2 \dot{\phi}_n^2\Big) \nonumber \\
        &-& D \sum_n  \Big(\exp[-a (r_n - R_0)]-1\Big)^2 \nonumber \\
        &-& K \sum_n (l_{n,n-1} - \ell_0)^2                  \nonumber \\
        &-& S \sum_n (r_n - r_{n-1})^2 \exp[-b (r_n + r_{n-1} -2R_0)]\,
\end{eqnarray}
where $\ell_0 =\sqrt{h^2 + 4 R_0^2 \sin^2(\theta/2)}$ and $l_{n,n-1}$ are
respectively the equilibrium and the actual distance between the two
bases $n$ and $n-1$ along a strand, 
\begin{equation}
l_{n,n-1} = \sqrt{{h}^2 + {r}_{n-1}^2 + {r}_{n}^2 - 2 {r}_{n-1} {r}_{n} \cos
({\phi}_{n} - {\phi}_{n-1})} \quad.
\end{equation}
For later purposes, it is convenient to introduce the local twist angle
defined as $\theta_n = {\phi}_{n} - {\phi}_{n-1}$.

The first term in the Lagrangian is the kinetic energy. The second
term is intended to describe the hydrogen bond interaction between the
two bases in a pair. Following Refs.~\cite{Pey89} and
\cite{Bar99a}, a simple Morse potential form is chosen. The
quadratic term in $(l_{n,n-1} - \ell_0)^2$ represents the elastic energy of
the backbone rods between neighboring base-pairs on each
strand. Finally, the last term models a stacking interaction between
neighboring base pairs. Its effect is to decrease the stiffness of the
open parts of the chain relatively to the closed ones and to stabilize
the latter with respect to the denaturation of a single
base-pair. Terms of this type increase the cooperative effects close
to the melting transition \cite{Dau95,Barbi}.

In the present paper we restrict our attention to a chain with free
boundary conditions, which corresponds to the experimental situation
when DNA denaturation is studied in solution. However, the model
described above can be easily modified to account for an external
torque $\Gamma$ applied at the base pairs at the two ends \cite{Coc99b}, as
it is done in some single-molecule experiments.

The geometrical parameters of the model can be straightforwardly fixed
according to the available structural data \cite{Bar99a}. Much more delicate
is instead the choice of the parameters $b,D,S$ and $K$ gauging the
effective forces.  We selected values similar to those previously considered
for the fixed-planes case, which have been discussed elsewhere: the choice
of the $K$ parameter can be independently derived \cite{Barbi} from the
twist persistence length \cite{Str96}, while the choice of the other two
parameters was based \cite{Cocco} on a comparison with recent mechanical
denaturation experiments  \cite{Str99}. In particular, the important
parameter $D$ which sets the main energy scale, has been tuned to reproduce
as closely as possible the experimental value of the denaturation
temperature $T_D=350 \,K$.  The full set of parameters used in the following
are summarized in Table~\ref{parameters}.

\begin{table}[h]
\begin{tabular}{lcrr}
\hline 
\hline 
Parameter & Symbol & Value\\
\hline
Morse potential range& $a$   & 6.3 & \AA $^{-1}$\\
Stacking interaction range& $b$   &  0.5 & \AA$^{-1}$\\
Morse potential depth& $D$  & 0.15 & eV\\
Stacking interaction coupling& $S$ & 0.65 & eV\AA$^{-2}$\\
Inter-plane distance& $h$  & 3.4 & \AA\\
Elastic coupling&  $K$  & 0.04 & eV\AA$^{-2}$ \\
Equilibrium distance& $R_0$ & 10  & \AA\\
Twist angle & $\theta$ & 0.60707 & rad\\
Base masses & $m$ & 300 & a.m.u.\\
\hline
\hline
\end{tabular}
\caption{The parameter set used throughout the paper}
\label{parameters}
\end{table}

For the numerics it is convenient to work in dimensionless units. A
suitable choice is to measure lengths and energies in the natural
units of the Morse potential, $a^{-1}$ and $D$ respectively, whereby
time is expressed in units of $\sqrt{m/Da^2}$. With the parameters
of Table~I, one time unit (t.u.) $\simeq 2.3$~ps.
 
\section{The denaturation transition: a qualitative discussion}
\label{sec:qualitative}

Before going on, it is instructive to briefly discuss the
thermodynamic state variables of the chain as well as the (possible)
analogies between denaturation and the more familiar liquid-gas
transition.

As already pointed out \cite{Coc99b}, the applied torque $\Gamma$
plays the role of the pressure $P$ for the liquid-gas system. Its
conjugate variable is the {\em degree of supercoiling} $\sigma$. Since
we do not consider the curvature of the axis of the helix, it
reduces simply to the {\em average twist}:
\begin{equation}
\sigma \;=\;  \sum_n^N \frac{(\langle\theta_n \rangle- \theta)}{N\theta}\,.
\end{equation}
This variable thus plays the role of the volume $V$. Following this analogy 
we can therefore establish the following correspondences:
\begin{center}
\begin{tabular}{ccc}
DNA model & & liquid-gas \\
$ T $ & $\longleftrightarrow$ & $T$ \\
$\Gamma $ & $\longleftrightarrow$ & $P$ \\
$-\sigma$ & $\longleftrightarrow$ & $V$ \\
\end{tabular}
\end{center}
The sign of $\sigma$ is chosen for convenience as the 
degree of supercoiling vanishes
for B-DNA and is negative for a partially denaturated chain (see also
Eq.~\ref{eq:minus} below).

Two natural scenarios may thus be
expected: an isothermal, torque-induced denaturation at
$\Gamma=\Gamma_D$ or a fixed-torque, thermally-induced one at $T=T_D$.
For the liquid-gas case, these two situations correspond to crossing
of the coexistence curve in the $(P,T)$ plane with an horizontal or a
vertical line, leading to the transitions classically described by
isotherms in the $(P,V)$ plane or by isobars in the $(V,T)$ one,
respectively. In both cases, the presence of a constant-temperature
or a constant-pressure/torque domain
is associated to the phase coexistence.

\begin{figure}[uh]
\includegraphics[width=80mm]{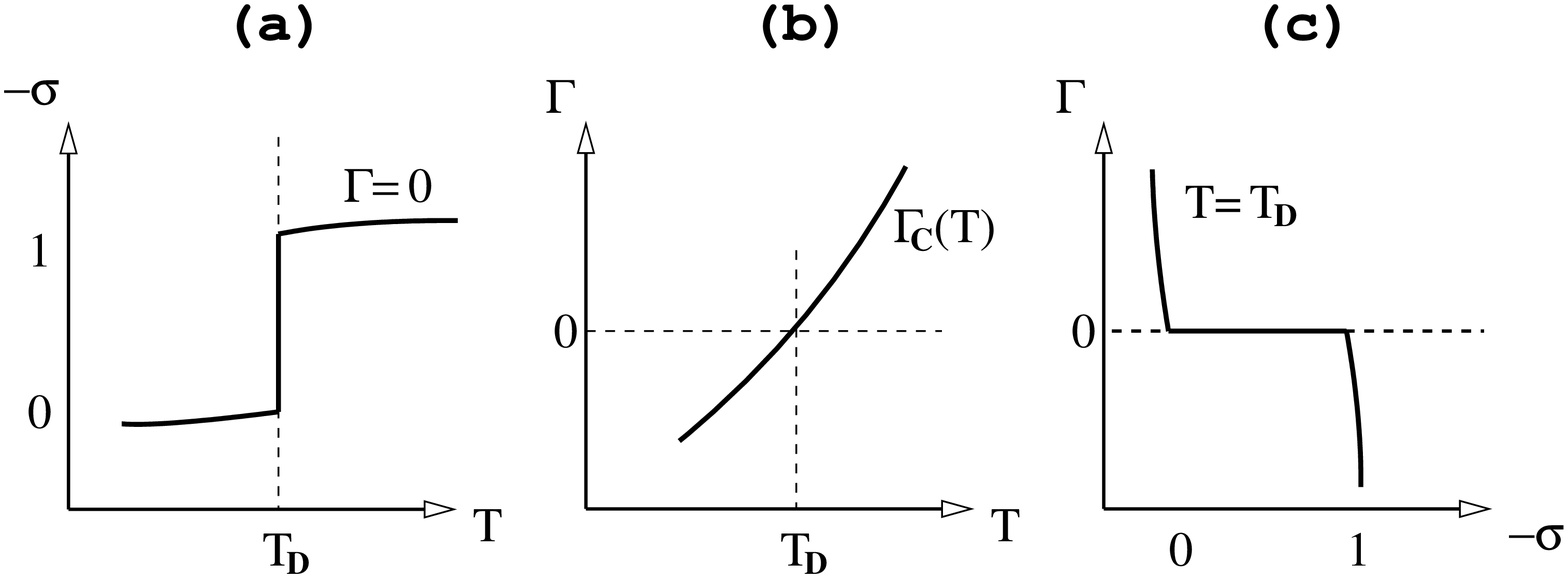}
\caption{A sketch of the behavior of first-order transition curves 
  in the three planes defined by the DNA model parameters $-\sigma$,
  $\Gamma$ and $T$.  $\Gamma_D$ and $T_D$ represent the transition
  torque and temperature, respectively;  (a) thermally-induced
  denaturation ``isobar'' at zero torque; (b) coexistence
  curve; (c) torque-induced denaturation isotherm at $T=T_D$,
  corresponding to a zero critical torque.  }
\label{fig:sketch}
\end{figure}

Within this analogy, the transition isotherms correspond to the curves
in Fig.~3 of Ref.~\onlinecite{Coc99b}, and reproduced schematically, on
the $(\Gamma,-\sigma)$ plane, in Fig.~\ref{fig:sketch}c. Conversely,
the thermal denaturation at constant torque $\Gamma=0$ are correctly
described in the $(-\sigma,T)$ plane (Fig.~\ref{fig:sketch}a).  Notice
that, with the convention that a negative $\Gamma$ corresponds to an
untwisting torque, both $\Gamma_D$ and $T_D$ must increase upon
increasing temperature and torque respectively (see 
Fig.~\ref{fig:sketch}b). However the
negative torque cannot exceed some critical value 
without leading to 
an instability of the helix associated to a change of the sign of the
helicity. In the following, we will focus on the thermal
denaturation transition at $\Gamma=0$.

It is important to remark that the helical constraints included in the
model roughly impose, at vanishing external torque, $\theta_n\approx
\theta$ for a closed chain segment, and $\theta_n \approx 0$ for the
denatured one. This follows from the geometry of the helix and
the stiffness of the strands: since the
distance between consecutive bases is constrained by the elastic
rods to be approximately equal to $\ell_0$, $\theta_n$ is of order
$\ell_0/r_n$ and hence very small for $ r_n\gg R_0$.  
Let us denote by $n_d$ the average number of open bases in a chain of
length $N$. Provided that open and closed
regions coexist along the helix, and that they are spatially well
separated (this is well confirmed by simulations as we show below),
then, to a good approximation, we have
\begin{equation} 
\label{eq:minus}
-\sigma \approx -\frac{n_d (-\theta)}{N \theta} = \frac{n_d}{N}  \; . 
\end{equation} 
The latter quantity in nothing but the average fraction of open base
pairs $\rho = n_d/N$ and the ``isobar'' $-\sigma(T)$ can thus be identified
with the familiar denaturation curve $\rho(T)$. In other words, the
supercoiling and the fraction of open base pairs are equivalent order
parameters. 

Before going further there is one crucial issue that should be
addressed. One may argue that for a one-dimensional model like the one
at hand no phase transition of any type should be observed. However,
all usual arguments against the existence of singularities in
thermodynamic potentials have been showed {\em not to hold} for the PB
model \cite{Dau02}. Since the latter is in many respects similar to the
helicoidal model, the same arguments apply and a genuine phase
transition is not forbidden a priori. This is confirmed by the
transfer integral approach which can be carried exactly (although
partly numerically) for this simple model.

\section{Transfer integral approach}  
\label{sec:transferint}
\subsection{\lq\lq Apparent" thermodynamics}
Because the model is one dimensional, a 
direct calculation of the partition function can be performed
by the transfer integral (TI) method, as it was done for the simpler PB
model \cite{Dau95}. The calculation proceeds along the same lines, but
it is more involved because the model has two degrees of freedom
($r_n$ and $\phi_n$) per unit cell. It nevertheless reduces to a 
one-dimensional TI equation because the contribution
introduced by the angular part can be diagonalized by a Fourier transform
\cite{Coc99b,Cocco}.  
As the calculation has already been presented in
these earlier studies we do not discuss the method here and we confine
our attention on its results, which point out new aspects of the
transition that had been overlooked.


In this subsection 
we restrict ourselves to results obtained via numerical
solution of the TI equation. The accuracy of this approach is limited first by
discretization errors in the integrations and by the need to numerically
evaluate integrals over an infinite domain. As discussed in the next
subsection, this second restriction can be partially lifted by a finite size
scaling analysis, which involves a properly controlled approach to
infinity. Integration methods, such as the Gauss-Legendre quadrature which
select appropriate absissa for the evaluation of the
function according to the number of points
involved in the calculation, are also useful to integrate over a large domain
with a reasonable number of points.  For this first study, in order to ensure
that all integrals are evaluated with the same discretization error, we have
computed them with a $10^{th}$ order Bode's method \cite{Abramowitz} with a
fixed
spatial step $\delta r = 0.032$~\AA\ and a minimum value $r_{min} =
9.7$~\AA\ (due to the strong repulsion between bases described by the
Morse potential, $r$ cannot take values significantly below the
equilibrium length of $10$~\AA). 
The maximum $r_{max}$ of the integration range depends on the number of
integration points, which has been varied from 631 to 3601 
leading to $29.9 \leq
r_{max} \leq 124.9$~\AA. The eigenvalues of the transfer integral
operator have been obtained either by diagonalization of the
equivalent matrix problem or by the Kellog's method \cite{Courant}
to get the two lowest eigenvalues.

The eigenvalues $\Lambda$ of the TI operator will
henceforth be written as $\Lambda = \exp (- \epsilon / k_B T)$ where
$k_B$ is the Boltzmann constant. With this notation the free energy
per particle is determined by the smallest $\epsilon$-eigenvalue
$\epsilon_0$ and is given by
\begin{equation}
\label{eq:freenergy}
f = - k_B T \ln( 4 \pi m k_B T) + \epsilon_0 \; .
\end{equation}
Relevant thermodynamic quantities
like the entropy and specific heat are then evaluated from the standard
relations 
\begin{equation}
s =- \frac{\partial f}{\partial T}, \qquad c_v 
  = -T \frac{\partial^2 f}{\partial T^2}  \quad; 
\end{equation}
 the mean base-pair
stretching is given by 
\begin{equation}
  \langle r \rangle = \int_{0}^{+\infty} r  |\phi_0 (r) |^2 \, dr\; ,
\end{equation}
where $\phi_0$ is the eigenfunction associated with
$\epsilon _{0}$.

\begin{figure}[h!]
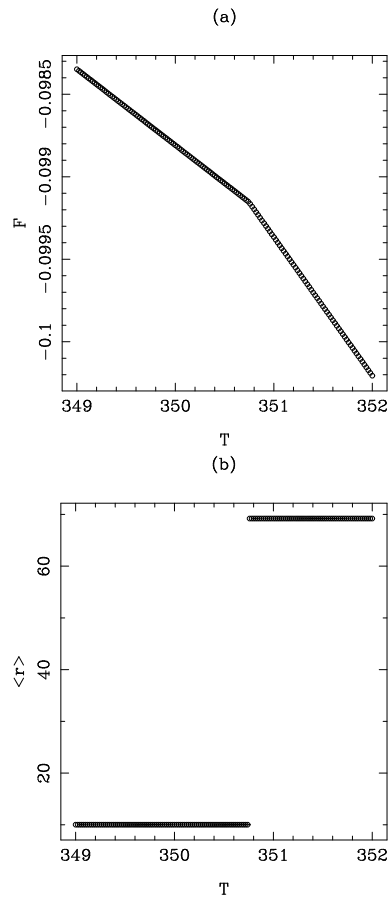

\begin{tabular}{c}
    \includegraphics[width=50mm]{fren1121.eps} \\
    \includegraphics[width=50mm]{rmoy1121.eps} \\
\end{tabular}
    \caption{The free energy per particle $f$ (a), and
    the mean value of the base pair stretching $\langle r \rangle$ (b) 
    of the model evaluated by the
    transfer integral method in the temperature range 
    $349 \le T \le 352$~K with a step $\Delta T = 0.02$~K. 
    The model parameters are those
    listed in Table~\ref{parameters}
    (Calculation with the Bode's method, $\delta r = 0.032$~\AA,
    $r_{min} = 9.7$~\AA, $r_{max} = 124.9$~\AA).
\label{fig:tiresults} }
\end{figure}

The free energy $f$ and the mean value of the base pair stretching $\langle r
\rangle$ for our model are shown in Figure~\ref{fig:tiresults}, for
temperatures going from 349 to 352~K with a step of 0.02~K.
Within the accuracy of the calculation, a cusp in $f$ at $T_D=350.74$~K is
distinctly seen.  It is associated with a sharp jump of the entropy at the
transition.  A jump in the specific heat is also observed.  Evaluating
numerically the first and second left- and right- derivatives of the free
energy, one obtains the jump in entropy $\Delta s=4.40~k_B$, or $8.75$
cal/K/mol, and the jump in specific heat $\Delta c_v=0.64~k_B$.  The specific
heat drops from $2.14 \, k_B$ below $T_D$  to $c_v = 1.5\,
k_B$ for $T > T_D$ as expected from equipartition because after denaturation
only the harmonic contributions of the hamiltonian stay significant.

Fig.~\ref{fig:tiresults}b shows that, within numerical accuracy,
$ \langle r \rangle $ exhibits a discontinuous 
transition from a finite constant value (very close to the equilibrium
value $R_0$) to a value of the order of the system size; in other words,
the eigenfunction $\phi _0$ appears to become suddenly delocalized.
The picture of a sharp transition persists down to a temperature
sampling of 0.01 K. 

Although the numerical results strongly suggest the occurrence of
a first order transition, caution is necessary: previous studies
of the related PB model have shown that
the nonlinear stacking produces
an extremely sharp, first-order-like behavior
which masks the underlying second-order transition
as long as one stays out of a
very narrow domain in
the immediate vicinity of the critical temperature $T_D$ 
(exponential crossover \cite{TheoToy,TheodorMadrid}).
A more complete picture of the properties of the transition
will therefore be given in what follows.

\subsection{The \lq\lq underlying" transition}

We first address the question of what happens in  the absence
of the nonlinear stacking, i.e. at $S=0$. Preliminary numerical 
investigations suggest a smooth behavior, both
of the lowest eigenvalue $\epsilon _0$ and of the next-to-lowest, 
$\epsilon _1$, as functions of temperature; an "avoided crossing" between
them appears, with a small, but finite gap which has a minimum
at a certain temperature.  Before we proceed to
analyze the data obtained in detail, it is necessary to provide
some background and notation.  

The order of the phase
transition of the ideal system of unconstrained transverse spatial
extent  is  determined by the critical exponent 
$\nu$ which characterizes the gap
$  \Delta \epsilon \equiv \epsilon _1 - \epsilon _0 \propto
 (T_D-T) ^{\nu}  $
at temperatures below  $T_D$; a value $\nu=1$ implies
a cusp in the free energy and a discontinuous entropy; 
a value equal to 2 implies a discontinuity in the specific 
heat, i.e. a usual 2nd order transition, etc.

\begin{figure}[h!]
  \includegraphics[width=6cm]{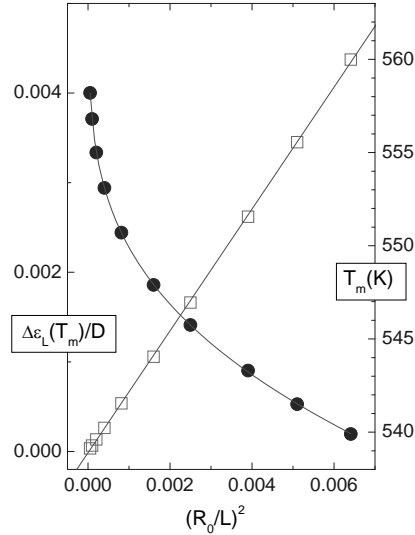}
\caption{Dependence of the minimal difference between the lowest eigenvalues
of the transfer integral operator $\Delta \epsilon_L(T_m)$ 
(open squares, left y-axis)
and temperatures $T_m$ (closed circles, right y-axis) versus $1/L^2$. 
}
\label{fig:TmL}
\end{figure}

The \lq\lq raw" data provided by numerical solution of the TI
equation refer to a particular transverse 
system size $L=r_{max}$ determined by
the imposition of an upper cutoff to the integration. On the other hand,
near a critical point of the infinite system, the transverse fluctuations of  
the order parameter also diverge. The quantity 
$\xi_{\perp} = \sqrt{ \langle r^2 \rangle - \langle r \rangle^2 }$
provides a measure of the distance from the critical point with
dimensions of length. According to the finite-size scaling hypothesis 
\cite{FSS_Fisher}, size-dependent properties of a system 
in the vicinity of the transition should
depend solely on the ratio $L/\xi_{\perp}$, i.e.
\begin{equation}
\Delta\epsilon_L (T)  =  L^{-\sigma} f_g \left( \frac{L}{\xi_{\perp} }   \right)
\quad,
\label{eq:fssgap}
\end{equation}
where the exponent $\sigma$  characterizes the rounding of
the gap (cf. below), $f_{g}(0)$ is a nonzero constant, and  
$f_g(x) \propto x^{\sigma }$ as $x\gg 1$  guarantees
size-independence in the limit $L \to \infty$.
In the simplest cases, the positions of the minima of the gap are related
to the type of divergence of $\xi_{\perp}$; according to the
above scaling scenario, the temperature $T_m(L)$ where the gap
minimum occurs, is such that
\begin{equation}
\xi_{\perp}\left(T_D - T_m(L)\right) \approx L \quad.
\label{eq:rounding}
\end{equation}
We have numerically solved the TI equation for a wide
range of system sizes, using Gauss-Legendre quadratures,
where $L$ is defined as the largest grid point value
provided by the Gauss-Legendre
algorithm for the number $N$ of grid points chosen for the calculation.
Again, in order to ensure uniform accuracy, $N$
grows proportionately to the system size, with
$r_{max}=350$~\AA\  corresponding to $N=2048$ points.
Fig.~\ref{fig:TmL}  shows the dependence of the minimal gap 
$\Delta \epsilon_L(T_m)$, and the corresponding temperature
$T_m$ on $L$. The gap appears to depend quadratically on $1/L$
(i.e. $\sigma=2$ in (\ref{eq:fssgap})~),
to very good accuracy, with an estimated limit
of $1.4~10^{-5}$ (with a standard deviation $0.8~10^{-5}$) 
as $L \to \infty $. 
We note in passing that the disappearance of the gap between
the bound state eigenvalue and the one belonging to the bottom of
the continuum band provides numerical evidence for the true occurrence
of an exact, thermodynamic phase transition \cite{TheodorScaling}. 
The temperatures which
correpond to the gap minima can be well fitted
to the function
\begin{equation}
T_m(L)  = T_D \left[ 1- \frac{a_2}{\ln\left(\frac{L}{R_0c}\right)} \right]
\label{eq:TmL}
\end{equation}
with  $T_D=577.8 K$, $a_2=0.170$ and $c=0.94$. 
This type of dependence
of $T_m$ on $L$ immediately suggests (cf. Eq.  \ref{eq:rounding}) that 
\begin{equation}
\xi_{\perp} = c R_0 e^{a_2/|\tau|}   \quad,
\label{eq:xiperp}
\end{equation}
where $\tau = T/T_D -1$.

\begin{figure}[h!]
\includegraphics[width=6cm]{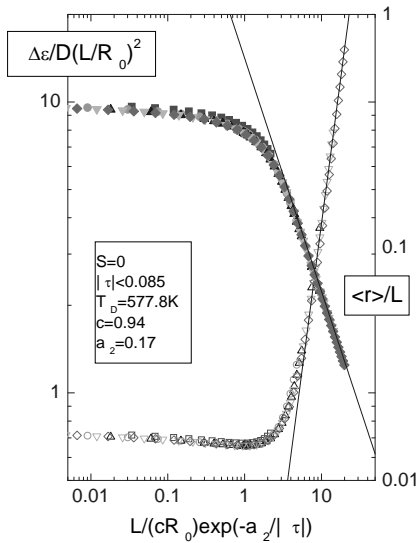}
\caption{Scaling of the difference between the two lowest eigenvalues
  of the transfer integral operator (open symbols, left y-scale), and
  the order parameter (solid symbols, right y-scale). 
  The different points have been
  obtained by transfer integral calculations performed with 5 values
  of $L/R_0 = 16,\, 35,\, 70,\, 100,\,
  140$ . The dotted lines have slopes $2$  and $-1$, respectively,
  in accordance with the 
  finite-size scaling hypothesis. 
}
\label{fig:diffscale}
\end{figure}

Fig.~\ref{fig:diffscale} shows that data taken from a wide  range of system sizes
scale well if plotted according to Eqs. (\ref{eq:fssgap}) and
(\ref{eq:xiperp}).   The numerical evidence thus strongly suggests that the
underlying transition manifests  itself as an essential singularity of the gap,
of the Kosterlitz-Thouless (KT) type. Eqs.\   (\ref{eq:xiperp}) and
(\ref{eq:fssgap}) then imply that, in the limit $L \to \infty$,
\begin{equation}
\Delta \epsilon  \propto e^{-2 a_2/|\tau|}   \quad.
\label{eq:LinfGap}
\end{equation}
In the Appendix, it will be possible to identify the origin of this
particular behavior as an inverse-square attractive interaction
between the streching coordinates of successive base pairs.

\subsection{Finite stacking revisited}

\begin{figure}[h!]
\includegraphics[width=6cm]{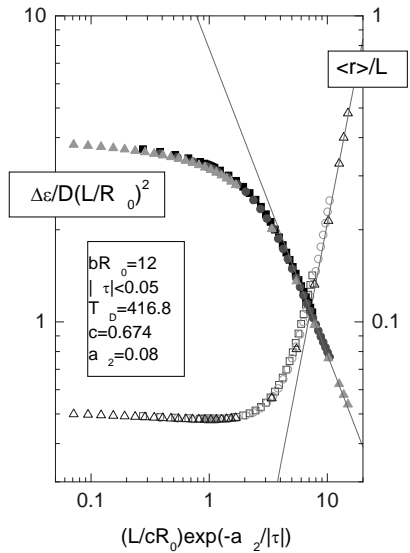}
\caption{Scaling of gap (open symbols, left y-scale), and
order parameter (solid symbols, right y-scale); model parameters 
are those of Table \protect\ref{parameters}, 
with the exception of $b=1.2$~\AA$^{-1}$.
The different points have been
  obtained by transfer integral calculations performed at
  $L/R_0= 25,\, 35,\, 50$. The dashed lines have slopes $2$  and $-1$, 
respectively,
in accordance with the 
finite-size scaling hypothesis. }
\label{fig:diffscale_1p2}
\end{figure}

\begin{figure}[h!]
\includegraphics[width=6cm]{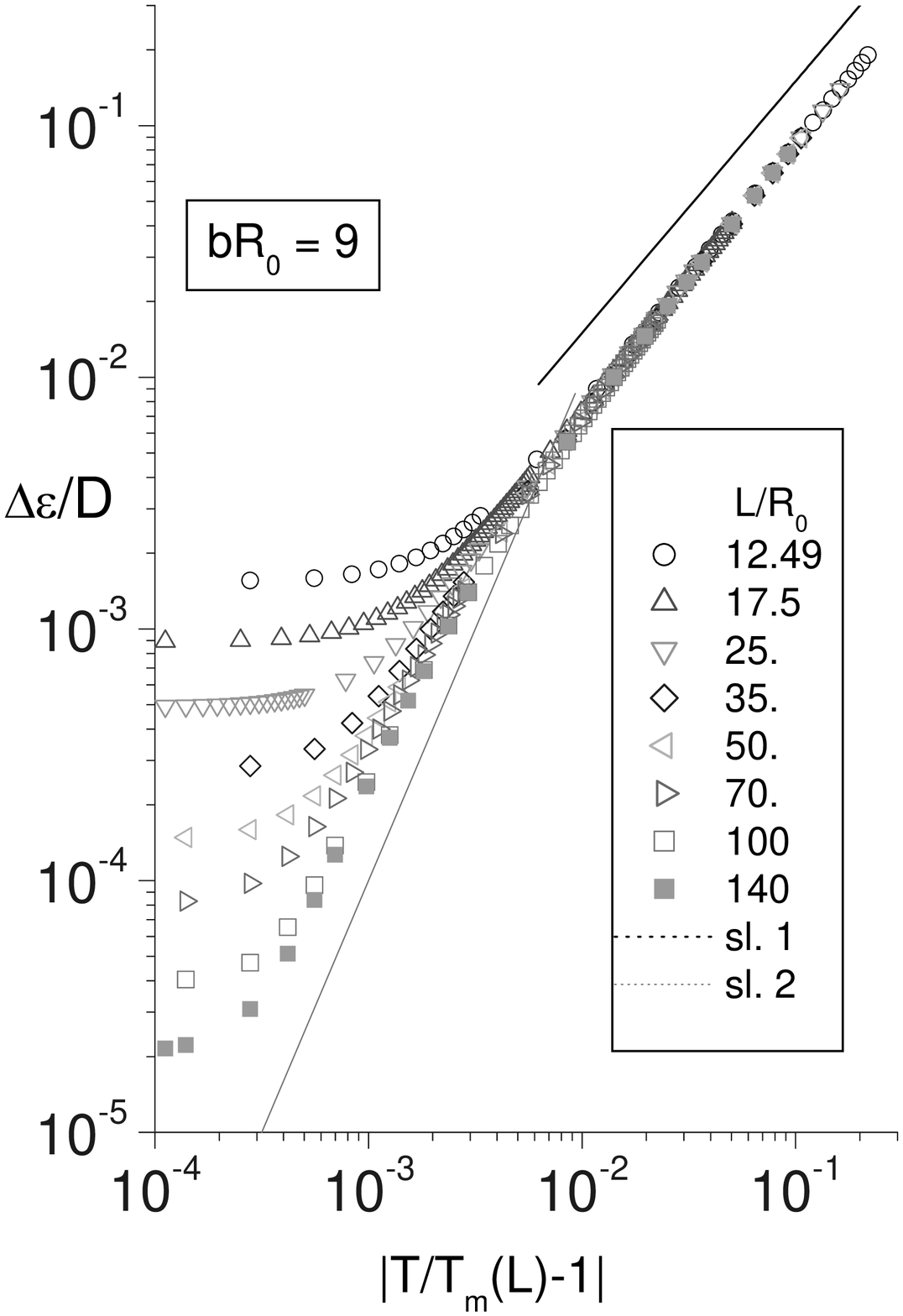}
\caption{Dependence of the gap 
on the reduced temperature $T/T_m - 1$ for $b=0.9$~\AA$^{-1}$
and system sizes $L/R_0 = 12,49,\, 17.5,\, 25,\, 35,\, 50,\, 70$.
The dashed and dotted lines have slopes $1$
and $2$, respectively. Apparent first-order
behavior prevails for $|\tau|>0.01$; closer to the
critical point there a clear increase in the slope,
before the onset of finite size rounding.
}
\label{fig:crossover}
\end{figure}

It is now reasonable to conjecture, by analogy with what happens
in the PB model, that the effects of the 
nonlinear stacking interaction will depend on its range. For the 
standard parameter set of Table \ref{parameters},
the ratio $b/a= 0.079$ is very
small indeed.  What happens at a less extreme regime, $b/a= 0.190$,
is shown in Fig.~\ref{fig:diffscale_1p2}.  Scaling according to the
Ansatz (\ref{eq:xiperp}) holds within a fairly narrow range $|\tau|<0.05$ 
around the denaturation point; note that it is the smaller
magnitude of the nonuniversal
parameter $a_2$ which is responsible for the narrowing of
the asymptotic critical region. 
Fig. ~\ref{fig:crossover} summarizes what 
happens at $b=0.9$~\AA$^{-1}$, i.e. $b/a=0.143$, only
slightly above the value of Table  \ref{parameters}.
The  gap exhibits an apparent  critical  exponent
very close to unity down to $\tau=0.01$; closer to
the denaturation point, the effective slope increases 
significantly;  it is reasonable to conjecture that 
at temperatures even closer to $T_D$, the
asymptotic  behavior will  be  dominated by  
the  underlying  essential singularity. 
At the physically relevant value of $b=0.5$~\AA$^{-1}$,
crossover to the KT regime has  moved below
$\tau=10^{-5}$  and is practically unobservable.\par

The analysis of this section demonstrates that, in spite of the very different
mathematical properties of their "bare" versions, both the "straight" (PB) and
the helicoidal DNA models, are effectively dominated by the stacking
interaction when the latter is of sufficiently long range; because of it, for all
practical purposes, the transition has all the characteristics of a first order
transition, including a practically infinite discontinuity of the mean base pair
stretching, and a latent heat.  Similarly, at the transition temperature, a very
small temperature gradient (of the order of the width of the transition region,
i.e.\ $\Delta T < 0.001$~K) leads to an apparent phase coexistence and hence to
many features that one would be tempted to qualify as ``typical of a first-order
transition'' as shown in the next sections.  These properties are very
reminiscent of some results found on models of martensitic phase transitions
\cite{Gooding,Gooding92,morris}, but because we are dealing with a
one-dimensional model that has a genuine phase transition, the phenomenon is more
remarkable here.

Our results provide an excellent example of the distinction between the
``experimental'' perspective and the ``theoretical'' one, regarding the
definition of the order of a phase transition: although, in theory, the
transition of the helicoidal DNA model is of infinite order (essential
singularity), the actual temperature range over which it manifests its
continuous character is far beyond the limits of either experimental or
numerical observation.

\section{Molecular dynamics}
\label{sec:md}

In this section we report the results of direct simulations of the
model. They bring complementary informations on the nature of the
transition and allow us to study its dynamics  as
discussed in Sec.~\ref{sec:dynamics}. As said above, we consider the
case of thermal denaturation for $\Gamma=0$ and free boundary
conditions.  Microcanonical and canonical simulations were performed
because they allow the observation of the phase space from different
view points.

In the microcanonical
ensemble, the Euler-Lagrange equations derived from (\ref{lagrangian})
were integrated directly with the standard fourth-order Runge-Kutta
method with a small enough time-step (typically $0.02$~t.u.) in order
to insure that the relative energy drift is negligible (usually
better than $10^{-5}$) on the time-scales of each run, i.e.\ 
$10^{5}$ to $10^{6}$~t.u.\ 
\cite{Symplectic}.  Initially, all particles are set in their
equilibrium positions ($r_n=R_0$, $\phi_n=n\theta$) with random
Gaussian distributed velocities (with zero average) in the radial
direction. The variance of the distribution serves to fix the energy
per degree of freedom $e$.  
The averaging of the quantities of interest
is only started after a long enough transient to let the
system equilibrate. After equilibration, the
thermal energy $k_B T$ is computed in the usual way as twice the
average kinetic energy per degree of freedom.

Constant-temperature (canonical) results were obtained through
an extended Nos{\'e}-Hoover method using a  
thermostat chain  \cite{Mar92} which is
specifically designed to constrain the total kinetic energy to
fluctuate around $N k_B T$, insuring at the same time the correct
(canonical) distribution of its fluctuations. A chain of 3
thermostats was employed with the first thermostat typical frequency equal
to the highest phonon frequency of the lattice, $\omega_M=\{a^2
  D+ 2 K (R_0 (1-\cos\theta)/\ell_0)^2]/m \}^{1/2}$. The integration of the
corresponding equations of motion was again performed with a
fourth-order Runge-Kutta scheme with typical time step of
$0.01$~t.u., and thermalization is achieved by a long-enough transient.
Changes in temperature were performed in a sequential way upon heating
the chain with a temperature ramp and relaxing afterward.

\begin{figure}[uh]
\begin{center}

\vspace{0.5cm}
\includegraphics[width=60mm]{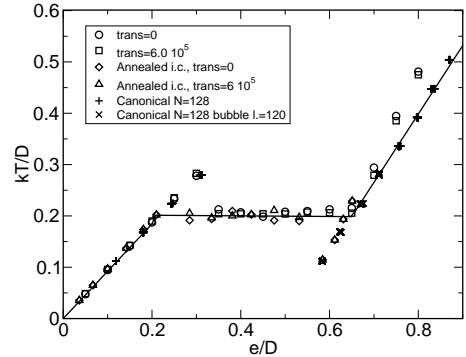}
\caption{Result of microcanonical simulations (open symbols): 
kinetic temperature 
as a function of the energy per degree of freedom for a molecule
of $N= 128$ base pairs. To show the convergence of averages, data for two 
transient durations are reported. Pluses and crosses respectively 
refer to canonical 
simulations with closed initial conditions and with the initial 
insertion of an artificial bubble of length $120$ base pairs 
(transient $10^5$ time units). 
Solid line is the transfer integral result (see text for details). 
}
\label{fig:micro}
\end{center}
\end{figure}

As discussed in Sec.~\ref{sec:transferint}, strictly speaking the
transition is smooth. However the temperature range of the 
crossover region to a smooth behavior is so small (less than
$\Delta T = 0.001$~K  for a stacking parameter $b=0.5$~\AA$^{-1}$) that
the numerical experiments, as well as actual denaturation experiments
on DNA, show all the character of a first order transition. Therefore
in this section we shall use the language of first order transitions
which is the appropriate language to discuss the results.

Measured caloric curves showing the temperature (in energy units) as a
function of the energy per degree of freedom, $k_BT(e)$, for a chain
of $N=128$ base pairs are reported in Fig.~\ref{fig:micro}. They
distinctly show a flat part at $k_BT=0.2012\, D$ corresponding to the
temperature $T_D=350$~K that has been found to be the denaturation
transition by the transfer integral calculation. In analogy with a
liquid-gas transition, the flat region occurring between $e_B=0.20\,D$
and $e_D=0.64\, D$ is thus identified as the curve of coexistence
between the closed and the denatured phases. The slope of the two
branches, $k_B T/e$ should be equal to $2/c_v$ (the factor 2 appears
because $e$, energy per degree of freedom is $\frac{1}{2}$ of the
energy per unit cell). The figure shows that this is in good agreement
with the values given by the TI calculation, i.e.\ $c_v \simeq 2.2$ below
$T_D$ and $1.5$ above $T_D$.
We have also compared the melting entropy per particle, 
$\Delta s=-\Delta(\partial f /\partial T)$, as obtained from the 
transfer integral calculation $\Delta s_{\rm th.}$, with that 
obtained from the microcanonical simulations 
$\Delta s_{\rm num.}$ for the finite chain i.e.\ 
the ratio $2(e_D-e_B)/T_D$:
\begin{equation} 
\Delta s_{\rm num.} = 3.70 \,  10^{-4} {\rm eV/K} ,\; \;
\Delta s_{\rm th.}  = 3.80 \,  10^{-4} {\rm eV/K} \,.
\end{equation}
The two quantities are in very good agreement.

In addition, the transition markedly displays a signature of {\em
metastability and hysteretic effects}. Indeed, the B-DNA branch
extends well above $T_D$ (up to about 500~K). Marked hysteretic
effect upon heating are also observed for the thermostated chain
(crosses in Fig.~\ref{fig:micro}).  Either in microcanonical and
canonical simulations, the system appears to be spontaneously
``trapped'' into this metastable state for low enough energies (or
temperatures) over the transition one.  Direct inspection of the
system configuration reveals that the chain is completely closed and
we can refer to it as an overheated state.

An undercooled branch exists as well below the denaturation temperature.  To
detect it in the microcanonical scheme, we employed the following procedure. The
initial condition, in the B-phase, is evolved for a certain time after which the
chain is ``annealed" by multiplying all velocities by an assigned factor smaller
than $1$ (we set it equal to $0.8$). The averages are thus computed after a
further transient (see again Fig.~\ref{fig:micro}). Similar results can be
obtained for the thermostatted chain by artificially imposing on standard initial
conditions the presence of a denaturated bubble of given length $\ell$ in the
middle of the chain at temperature $T > T_D$. We will give more details on this
procedure at the end of this section.

Canonical and microcanonical results are therefore consistent, apart
from some deviations at high temperatures, for $T > T_D$, which can be
expected because after denaturation the model becomes almost purely
harmonic as the Morse potential linking the bases plays no role for
base-pair distance $r$ corresponding to the plateau of the
potential, and the stacking contribution also tends to
vanish. Therefore for $T > T_D$ a microcanonical equilibrium cannot be
achieved, unless we force it by averaging over thermalized initial
conditions (for instance obtained by a Monte-Carlo procedure) or by a
temporary switch to a canonical simulation during a run.

The results were checked to be robust with respect to the transient
duration  as well as to the rate at which temperature is changed
through the ramp.  Alternative thermalization schemes do not change the
outcomes as well.  For example, simulations where microcanonical runs
are alternated to canonical  ones, yield the same results (except at
$T>T_D$ as mentioned above).  In such a case the computed averages are 
microcanonical as the thermostatted dynamics only serves as a way to
change  the system energy. 

Obviously, a crucial issue is the dependence of the results on the chain size.
We observed that upon increasing the chain length up to  
$N=256$ or $512$, the only
difference with respect to the $N=128$ case is a slower convergence
of the averages  in the high and intermediate energy region. Nonetheless, the
coexistence line is practically reached within comparable simulation times.  
This is presumably influenced by the initial conditions and could be improved 
by a more sensible choice.

\begin{figure}[uh]
\begin{center}
\includegraphics[clip,width=80mm]{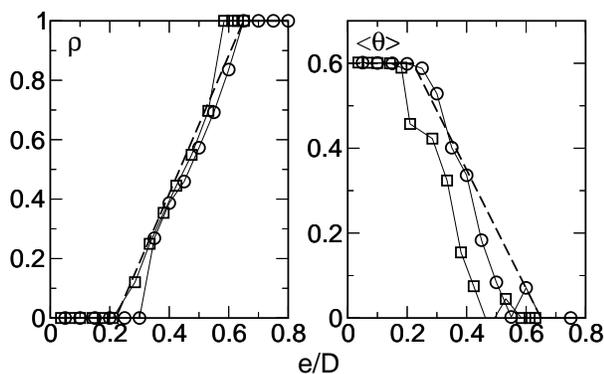}
\caption{ 
Fraction of open  base pairs $\rho$ and average twist $\langle \theta_n
\rangle$ from  microcanonical simulations.  The simulation parameters
are the same as in Fig.~\ref{fig:micro}. The circles were obtained with
an initially closed chain while the squares refer to annealed initial
conditions. }
\label{fig:lev}
\end{center}
\end{figure}

In order to precise the nature of the transition from a view point
close to experiments, we measured the average fraction of open base
pairs $\rho$.  This is indeed a quantity which is measurable by UV
absorption. As in previous studies \cite{Dau95,Bar99a,Coc99b}, we
consider a base pair to be open whenever the radial displacement $r_n$
is larger than the inflection point of the Morse potential i.e.\ when
$r_n/R_0> 8\ln2 $, and we average the counting during the run. A
simple reasoning shows that the order parameter $\rho$ should obey a
''lever rule"
\cite{Callen}
\begin{equation}
\label{lever}
e \; = \; (1-\rho) e_B \, + \, \rho e_D ,\qquad e_B < e < e_D
\end{equation}
thus implying that $\rho(e)$ increases linearly between 0 and 1 along
the coexistence line. In a similar way, we expect
that the average twist per base pair $\langle \theta_n \rangle
=\langle \phi_N-\phi_1 \rangle/N = \theta (1+\sigma)$ should decrease
linearly from a value close to $\theta$ to 0. This is illustrated in
Fig.~\ref{fig:lev}.  Notice once again the hysteretic effects.

\begin{figure}[uh]
\begin{center}
\includegraphics[clip,width=55mm]{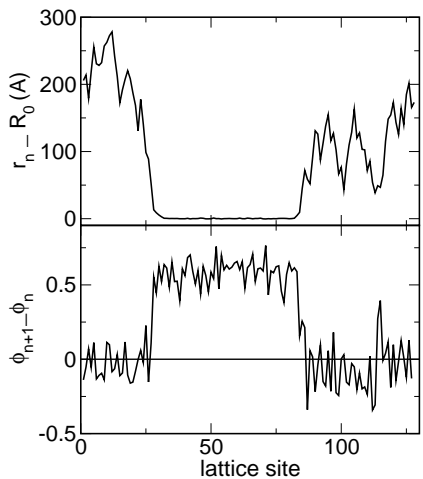} 
\caption{ 
Snapshot of the chain of 128 bps in the coexistence region 
$e=0.5 \,D$. 
}
\label{snaps}  
\end{center}
\end{figure}

From Fig.~\ref{fig:micro} it is clear that the microcanonical ensemble has the
merit of allowing to investigate the dynamics of the chain in the coexistence
region. For illustration, Fig.~\ref{snaps} shows a snapshot of the state of the
chain for $e/D=0.5$ were, from formula (\ref{lever}), we expect around 40\% of
the base pairs to be denaturated. The fact that the chain opens at the sides is
clearly caused by the free boundary conditions. Moreover this figure can give a
hint on why we observe a transition that has all the features of a true first
order transition (coexistence of phases, metastability) although the transition
is actually second order.  From a theoretical point of view, the order is
determined in the thermodynamic limit, i.e.\ for an infinite system. This
correspond to the identification, in the transfer integral method, of the free
energy with the lowest eigenvalue $\epsilon_0$, Eq.~(\ref{eq:freenergy}). In
numerical simulations, as well as in experiments, one is dealing with a finite
system. When the thermodynamic transition is extremely sharp (as it is the case
for the stacking parameter $b=0.5$~\AA$^{-1}$), the inhomogeneity caused by the
free ends is sufficient to lead to an apparent coexistence of phases, i.e.\ a
first-order-like transition, presumably because the boundary effects induce a
perturbation (in particular on the average local torque) which is sufficient to
change the local transition temperature by the very small amount which
separates the domain of closed DNA from the domain where the molecule
denaturates. This is why the molecular dynamics simulations are useful to
complete the transfer integral study, and to provide results that can be
compared with experiments.

\begin{figure}[uh]
\begin{center}

\includegraphics[clip,height=55mm]{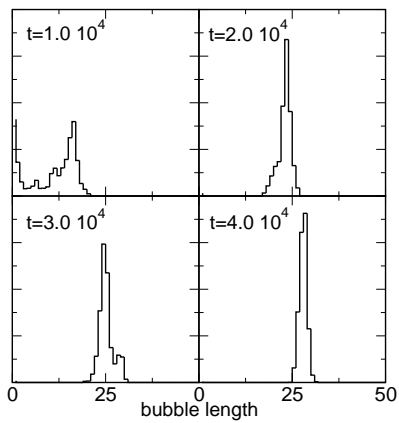}
\caption{ 
Distributions of the lengths of the denaturated regions at different
times in the coexistence region  $e=0.5 \,D$. To improve the 
statistics, histograms were cumulated over a time window of around
$10^3$ time units. 
}
\label{histo}  
\end{center}
\end{figure}

Another interesting aspect which can be studied through simulation is
the dynamics of opening events. This allows to look for analogies with
the classical nucleation mechanisms that drives relaxation from
metastable states at ordinary first-order transitions
\cite{Callen}. Fig.~\ref{histo} reports the distribution of the length
of denaturated bubbles for subsequent times during the same run of
Fig.~\ref{snaps}. There is a clear tendency for smaller bubbles to
close (or merge) until only a few large ones remain. A similar measure
in the overheated metastable phase shows instead that the size of
bubbles is pretty small and decrease systematically in time.

To further investigate this aspect, we performed simulations in the
canonical ensemble, starting from a thermalized state at temperature
$T$ and artificially seeding a denaturated bubble of given length
$\ell$ in the middle of the chain. To accomplish this, given the
geometry of the model, we set $\theta_n=0$ in the central region and
impose a triangular profile for the $r_n$s designed in such a way that
the resulting stress on the backbone springs is approximatively
zero. The flanking regions are initially at equilibrium and free from
any additional supercoiling.

In a first series of simulations we checked that for $T<T_D$ the B-DNA
form is very robust with respect to such local perturbation: for
instance, at $T=200$ or $300$~K bubbles of length $\ell\leq 12$
base pairs on a chain $128$ base-pair long tend to shrink and the system
rapidly returns to its completely closed state. Very large bubbles
($\ell \sim 110$ base pairs) may tend to close as well, although on longer
times. However, several cases are found where perturbations as large
as $\ell = 120$ base pairs are able to drive the system into the
metastable, undercooled, denaturated state. These experiment allow the
observation of the second metastable state in the canonical scheme:
thus completing the correspondence with the microcanonical results
(see again data shown in Fig.~\ref{fig:micro}).

The situation is, as expected, the opposite in the overheated region 
$T_D < T \lesssim 600$~K. Here, the insertion of a short bubble suffices 
to destabilize the B-DNA form and let the system switch to its equilibrium 
state, i.e.\ the completely open chain. For instance, at $T=400$ or $500$~K 
a bubble of length $\ell>8$ on a chain of $128$ base pairs is generally 
enough. Approximated estimates seem to indicate that the minimal
length of the bubble tend to decrease with temperature, as
intuitively expected, but this behavior is not very
systematic. Statistics over a very large number of events would be
necessary to conclude quantitatively.

\section{Dynamical structure factors} 
\label{sec:dynamics}

One of the motivations for considering mechanical models is the possibility
to probe microscopic and collective motion in different phases. 
In this section, 
we focus on dynamical correlation functions that usually reflect  
different types of excitations. More precisely, we computed the 
radial structure factor
\begin{equation}
S_r(q,\omega) \;=\; \left\langle \left| \int \sum _n \, r_n 
\, e^{i(qn-\omega t)} \, dt\right|^2 \right\rangle
\end{equation}
and the angular one $S_\psi(q,\omega)$ where $\psi_n=\phi_n-n\theta$ 
is the angular displacement from the equilibrium position. Brackets 
denote an average over an ensemble of independent molecular dynamics
trajectories (typically hundreds). All the results reported in this section 
are obtained in the microcanonical ensemble.   

\begin{figure}[uh]
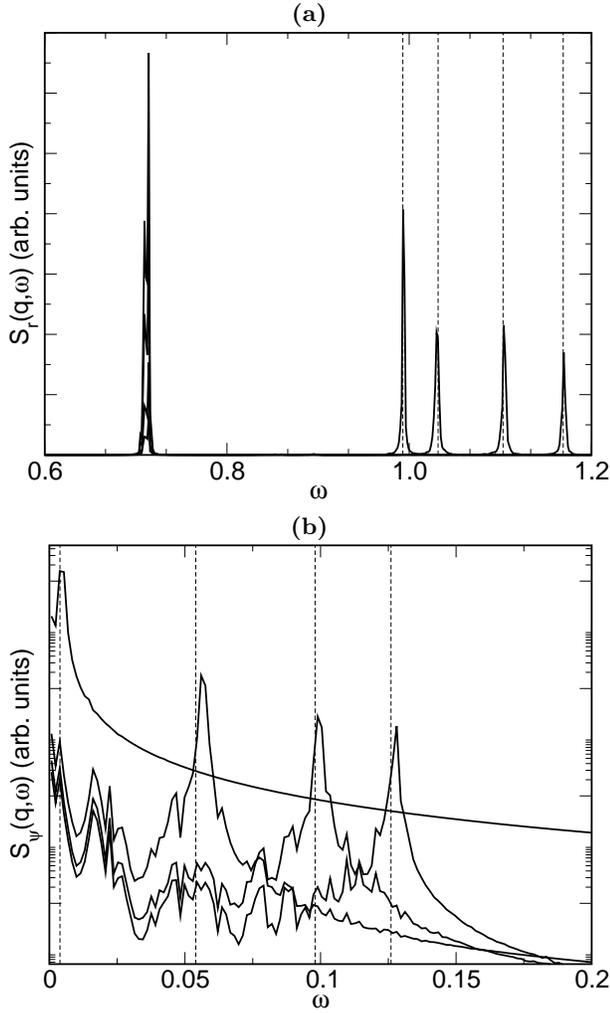

\begin{center}
\begin{tabular}{c} 
\textbf{(a)} \\
\includegraphics[clip,width=80mm]{sr-N256-0e01.eps} \\
\textbf{(b)} \\
\includegraphics[clip,width=80mm]{sp-N256-0e01.eps}
\end{tabular}
\caption{Structure factors $S_r(q,\omega)$ (a) and 
$S_\psi(q,\omega)$ (b) for $N=256$ at very low energy, $e=0.01\,D$
(corresponding to $T=18$~K). The different curves correspond to
wavenumbers $q= 2.5676,\, 1.66896,\, 0.88356,\, 0.09812$ (right to
left).
}
\label{sp}
\end{center}
\end{figure}

Let us first consider the low temperature native phase.  As shown in
Fig.~\ref{sp}, the spectral analysis display, as expected, sharp lines
at the frequencies of the two phonon branches $\omega_\pm(q)$ that
can be computed at $T=0$ in the harmonic approximation \cite{Barbi}
(see the vertical lines in Fig.~\ref{sp}).  Acoustic vibrations in the
angular variables are only weakly coupled to the radial (optical)
ones.  Interesting enough, the radial spectra also displays a large
peak at a frequency lying in the phonon gap and independent on the
wavenumber (i.e.\ the large peak at $\omega \simeq 0.7$ in
Fig.~\ref{sp}a. Its origin can be traced back to the excitation of a
localized surface mode. This is confirmed by direct inspection of the
chain configuration. Actually, the mode is found to slowly decay in
time due to nonlinear interaction leading to a systematic decrease of
its spectral component.

\begin{figure}[h!]
\begin{center}
\includegraphics[clip,width=85mm]{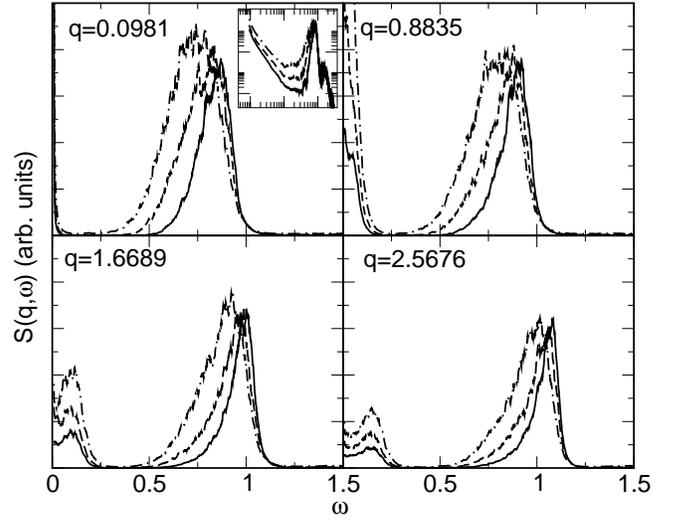} 
    \caption{Radial structure factors $S_r(q,\omega)$   
       for $N=256$ and different energies $e=0.17\,D$, $0.20\,D$, $0.30\,D$
	 (solid, dashed, dot--dashed lines respectively).   
       To reduce fluctuations, a 
       smoothing of the data has been performed by averaging over 10 
       consecutive channels.
        }
\label{sp2}
\end{center}
\end{figure}

Upon increasing the energy, the optical branch gradually shifts
towards lower values of the frequency (softening) and higher-harmonics
appear. Furthermore, the resonances in both the radial and angular
peaks are substantially broadened due to increasing anharmonicity that
enhances the effective damping. More importantly, a large
low-frequency component, a {\em central peak}, arises in the radial
structure function. The temperature dependence of $S_r(q,\omega)$
across the denaturation transition is illustrated in Fig.~\ref{sp2}.
The three different energies correspond to $T=300$, $357$ and $535$~K.
The latter value is well into the metastable overheated region. 
For fixed $q$, the position of the
central peak is unchanged upon increasing temperature but its width
broadens.  Furthermore, the $\omega^{-2}$ behavior at low frequencies
(see the inset of Fig.~\ref{sp2}) suggests a Lorentzian line-shape.
The origin of this central peak, also found in the simpler PB model,
and its properties are still unclear although it is tempting to assign
it to the slow dynamics of the bubble boundaries.

\begin{figure}[h!]
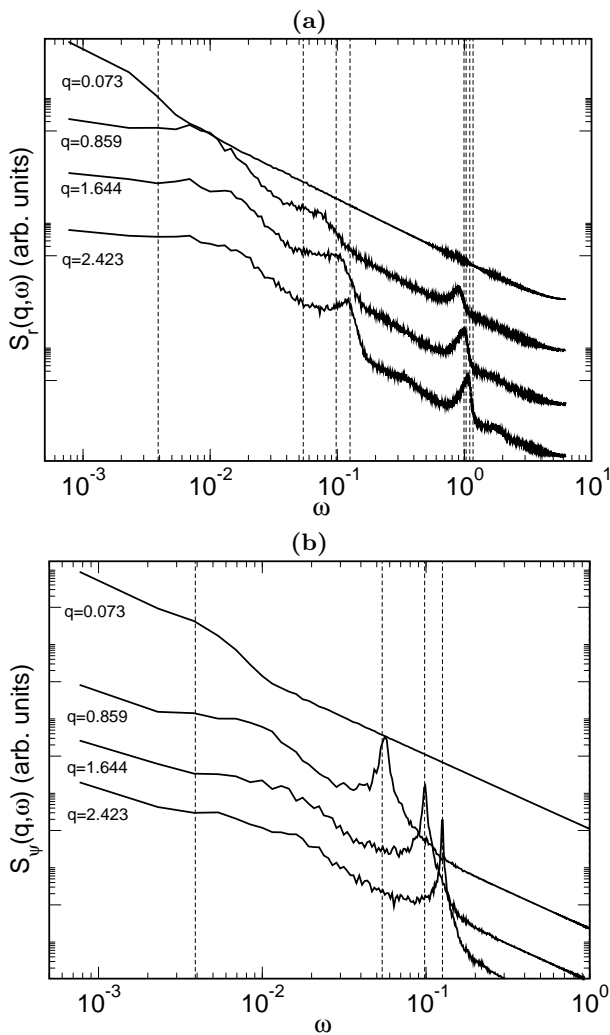

\begin{center}
\begin{tabular}{c}  
\textbf{(a)} \\
 \includegraphics[clip,width=80mm]{sr-N256-0e50.eps} \\
\textbf{(b)} \\
 \includegraphics[clip,width=80mm]{sp-N256-0e50.eps} 
\end{tabular}
    \caption{Structure factors $S_r(q,\omega)$ and 
     $S_\psi(q,\omega)$ for $N=256$ in the coexistence region
     $e=0.50\,D$. Vertical lines are the phonon frequencies (one
     from the acoustic and one from the optical branch for each 
     $q$ value) calculated  at $T=0$. Graphs have been arbitrarily 
     shifted for clarity.}
\label{spcoex}
\end{center}
\end{figure}

An even more sizable central component appears when closed and  open
form coexist (see Fig.~\ref{spcoex}). This is accompanied by a stronger
coupling between angular and radial degrees of freedom, as manifested
by the  peaks at acoustic frequencies in $S_r$.  The birth of large
low-frequency components bears  strong resemblance  with  heterophase
fluctuations observed in other lattice models  with (pseudo)
first-order transition characterized by large entropy barriers
\cite{Gooding,Gooding92,morris}. In other words, the  motion of the interface
between the two  phases should be responsible of the slow dynamics.

To close this section, it is worth mentioning that a related analysis of
collective modes for the helicoidal model has been recently reported 
\cite{Coc99c}. The analytical calculations were performed at  room-temperature
and are based on the instantaneous normal-modes.  At variance with our 
simulations, this approach describe the short-time dynamics (on a time scale
of picoseconds) and a direct comparison is therefore not straightforward.

\section{Conclusions and discussion}

The study of a simplified model of DNA has proved to be extremely fruitful to
unveil the basic features of the melting transition at the single--molecule
level. From the theoretical point of view, dealing with a one-dimensional model
(with two degrees of freedom per base pair) turned out to be particularly
convenient as it allows an exact evaluation of the partition function. Indeed,
the angular variables can be eliminated by Fourier transform, yielding a more
tractable one-variable transfer integral problem. The latter cannot  be solved
analytically but the numerical and approximate results presented above provide a
complete insight on the nature of the transition.  In particular, the finite-size
scaling analysis of the transfer integral turned out to be essential to take into
account the finiteness of the integration range. Such an analysis strongly
suggests that the underlying transition is continuous, of the Kosterlitz-Thouless
type. This behavior can be related to the existence of an effective attractive
force which is {\em directly connected to the helicoidal geometry} because it
appears when the angular degree of freedom is integrated out. Qualitatively it
can be understood as coming from the difficulty to disentangle the two helices.
On the other hand, for physically relevant values of the parameters, the
temperature range over which the continuous aspect of the transition can be
detected may become extremely narrow (less than $\Delta T =0.001$~K).  For all
practical purposes the transition appears to be perfectly sharp, and bears the
hallmarks of a first order transition, in agreement with experiments. In our
view, this result is remarkable and  attracts the attention on how numerical or
experimental observations on finite systems and with a limited resolution may
dramatically differ from theoretical expectations.

Molecular dynamics simulations confirm this apparent first-order character.
They show hysteresis and metastability as well as a coexistence region between
an open and a closed ``phase'' and, by varying the energy density in the
critical  region, a gradual change of the volume fraction occupied by the two
which is reminiscent of, say,  the liquid--gas transitions. 

In addition to the very sharp transition found by the theoretical analysis, the
finite--size effects certainly play a major role in the above phenomenology.
The helicoidal model is more sensitive to these finite size effects than the
``flat'' PB model because the free ends allow a release of the torsional energy
which appears when a segment of the chain opens. One can understand the crucial
role of boundary effects if one considers that a finite closed loop of
helicoidal DNA cannot denaturate at all because the two strands are entangled.

The dynamics of the transition, as probed by the calculation of the radial and
angular structure factors, shows some prominent features such as the existence
of a central peak that is presumably due to the slow motion of the denaturation
bubbles. Moreover, the coupling between opening and twist introduces some
additional spectral features that would deserve further investigations. 

Another point that should be reconsidered is the nucleation of denaturation
bubbles. The phenomenology described at the  end of Sec.~\ref{sec:md} is, at
least qualitatively, very much reminiscent of the nucleation mechanisms that
drives relaxation  from metastable states at ordinary first-order transitions
\cite{Callen}. For instance, simulation in the overheated state suggest the
existence of a  ``critical size" of the denaturation loops above which they
become unstable.  Hence, metastability stems from the fact that small enough
bubbles close relatively  fast. Nevertheless, there are important differences
that one should keep in mind.  Indeed, in classical nucleation theory the key
role is played by the surface  tension term (proportional to the square of the
droplet's radius),  whereby in our one-dimensional case the bubble ``surface''
is independent  of its length. The correspondence with the usual theory is
probably due to the torsional energy, associated to the opening, which grows
with the bubble size. This suggests that a ``one-dimensional nucleation
theory'' could be developed for helicoidal DNA.

The present study has focused on a DNA model that describes the molecule at the
scale of the base pair. We think that it is relevant because it is the scale of
the genetic code at which phenomena related to biological functions occur. The
helicoidal geometry itself is at this scale (or more precisely at the scale of a
few tens of base pairs). On the other hand, there are other phenomena that enter
in the statistics of DNA melting, and they are related to the behavior of the
molecule at a much larger scale, on which the strands are regarded as flexible
strings. Recent studies have shown that the entropy of the loops also
contribute to lead to a first order transition, provided that self avoiding
aspects betwen segments of the loop and  between open regions and closed  domains
are properly taken into account \cite{Muk00}. Our approach is complementary to
these studies and shows that the observed sharp melting transition of DNA may
have multiple origins.

\appendix
\section{Results obtained via gradient expansion}
\subsection{An approximate TI kernel}

\begin{figure}[h!]
\begin{center}
\includegraphics[clip,width=50mm]{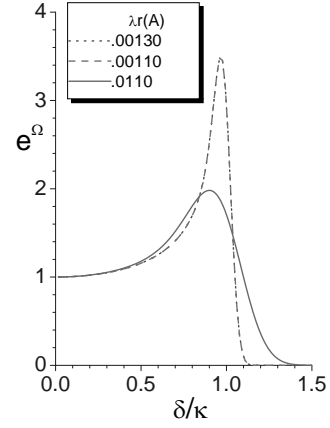} 
    \caption{ $\exp(\Omega )$ as a function of  $\delta /\kappa $
for two different values of the dimensionless ratio $\lambda $;
the $r$-dependence is not visible for $\lambda = 0.001$.
        }
\label{fig:OmApprox}
\end{center}
\end{figure}

In the absence of an external torque,
the nontrivial part of the partition function of the model is given  by
\begin{equation}
Z_{P}  = \int \prod_{n=1}^{N} \left\{dr_{n}\left[r_{n}r_{n-1}\right]^{1/2} 
d\theta _{n}
\right\} e^{-V\left( \{ r_{i},\theta _{i}  \} \right) /k_{B}T}  \quad,
\label{Ztot}
\end{equation}
where the potential energy consists of the three last 
terms in (\ref{lagrangian}), and 
the relative angle coordinate enters only via the second
term. Introducing sum and difference coordinates $
{\bar r_{n}}= (r_{n} + r _{n-1})/2$,  $
 \delta _{n}=  r_{n} - r _{n-1} $, 
it is possible to write the integral over $\theta _{n}$ as 
\begin{equation}
\int_{0}^{\pi } d\theta _{n}e^{- K (l_{n,n-1} - \ell_0)^2 / k_B T   } \equiv 
\Phi _{0}({\bar r_{n}}) \> e^{\Omega (\delta _{n}^{2 },{\bar r_{n}})}  \quad.
\label{inttheta}
\end{equation}
For the parameter values given in Table~\ref{parameters}, and $T=480$~K, the
dimensionless ratio $\lambda = k_B T / K R_{0}^{2 }$
is equal to $0.01$; this allows us to use the leading-order 
low-temperature asymptotic 
expansion
\begin{equation}
\Phi _{0}({\bar r}) \sim \sqrt{\frac{ \pi k_B T}{ K  }}
\frac{ L_{0}}{\kappa \bar r}
\left( 1 - \frac{ \kappa ^{2 }}{4{\bar r}^{2 } }\right)^{-1/2 }  \quad, 
\label{eq:Phi0}
\end{equation}
where $\kappa =2R_{0}\sin(\theta /2) =5.98$ ~\AA, over the whole temperature 
range of interest. Note that due to the repulsive core of the Morse
potential, the inequality 
\begin{equation}
r\gg \kappa /2
\label{eq:rineq}
\end{equation}
always holds. 
Furthermore, numerical evaluation reveals
that the function $\Omega $ is (i) almost independent
of ${\bar r}$, and (ii) weakly dependent on temperature 
(cf. Fig. \ref{fig:OmApprox}).
In the following we will use the temperature-independent 
approximation
\begin{equation}
\label{eq:OmApproxG}
e^{\Omega (\delta ^2)} \approx e^{-(\delta /\kappa )^{2}}
\quad,
\end{equation}
which misses the weak peak near $\delta = \kappa $,
but reproduces correctly the second moment, which is
central to what follows.
Within the approximations (\ref{eq:Phi0}), (\ref{eq:rineq}), 
the partition function is dominated,
in the thermodynamic limit, by the largest eigenvalue of the one-dimensional
TI equation
\begin{equation}
\int_{0}^{\infty}dr' T(r,r') \psi _{\nu }(r') = \Lambda _{\nu}\psi _{\nu }(r)
\label{1DTI}
\end{equation}
with 
\begin{eqnarray}
\nonumber
T(r,r') & \approx &  \sqrt{\frac{ \pi k_B T}{ K  }}\frac{ L_{0}}{\kappa}
\left( 1  + \frac{ \kappa ^{2 }}{8{\bar r}^{2 } }\right)
\left( 1  - \frac{ \delta  ^{2 }}{8{\bar r}^{2 } }\right)\\
&& \times e^{\Omega (\delta ^{2}) }e^{-V_{M}(r)/k_{B}T } 
e^{-V_{S}({\bar r}, \delta ^{2})/k_B T }
\label{ApprKern}
\end{eqnarray}
where $V_M(r) = D\left(1-\exp\left[-a\left(r-R_0\right)   \right]\right)^2$
and $V_S = S\delta ^2 \exp\left[-2b\left({\bar r}-R_0\right)   \right] $.

The form of $\Omega$ (cf. Fig.  \ref{fig:OmApprox}    and/or Eq.  
\ref{eq:OmApproxG} ) establishes that, in addition to the 
ranges $1/a$, $1/2b$ of the Morse and stacking interactions, respectively,
there is a third, much larger, characteristic length in the problem, $\kappa $.
Depending on the strength of the various parameters, it may be possible
to further simplify the general one-dimensional TI problem
and elucidate the ensuing critical behavior. 
Two distinct cases will be considered below.
\subsection{Strong stacking interaction: transformation to an ODE}
A gradient expansion of
(\ref{1DTI}) involves: (i) introducing
\begin{equation}
\psi _{\nu }(r+\delta ) \approx  \psi _{\nu }(r )  + 
\psi _{\nu }^{'}(r) \delta + 
\frac{ 1}{2 } \psi _{\nu }^{''}(r) \delta ^{2} 
\quad, 
\label{gradexp}
\end{equation}
(ii) changing the variable of integration from $r'$ ($=r + \delta $) 
to $\delta $, and (iii) performing the Gaussian integrals over $\delta $.
Noting that  (a) the combined effect of the stacking interaction 
and the Gaussian approximation
(\ref{eq:OmApproxG}) can be described in terms of the quantity
\begin{equation}
\frac{S}{k_{B}T}\mu ^{2}(r) \equiv 
\frac{S}{k_{B}T} e^{-2b\left({r}-R_0\right) }  
+\frac{1}{ \kappa ^{2}} \quad,
\label{eq:defmu}
\end{equation}
( note that $ S \mu^2$ can be interpreted as an effective nearest neighbor
harmonic spring constant )
and that (b) ${\bar r} \approx r $ to second order in $\delta $ everywhere in
(\ref{ApprKern}), one obtains
\begin{equation}
\left( 1 - \frac{k_{B}T }{16S\mu ^{2 }r^{2 } }\right) \psi _{\nu }
+\frac{k_{B}T }{4S\mu ^{2 } }\psi _{\nu }^{''}
= e^{-\beta \left( \epsilon _{ \nu } - U_1\right) }\psi _{\nu } 
\label{eq:intermODE}
\end{equation}
where $U_{1}(r)=V_M(r)+V_L(r)+V_B(r)$
and $\Lambda _{ \nu }= (\pi k_B T /\sqrt{KS})\ell_0/\kappa 
e^{-\beta \epsilon _{ \nu } }$; 
here, 
$V_L(r)= -k_B T (\kappa /r)^{2 }/8 $
is a long-range attraction which comes from in exponentiating
the  term in the first parentheses of (\ref{ApprKern}), and
$V_B(r)=k_B T \ln[ \mu (r)/\mu (\infty ) ]$ is a thermally 
generated barrier analogous to
the one described in \cite{The00} in the context of 
the one-dimensional DNA model with stacking. 
Expanding the exponential in the r.h.s. of (\ref{eq:intermODE})
and rearranging terms, one obtains
\begin{equation}
-\frac{(k_{B}T)^2 }{4S\mu ^{2 } }\psi _{\nu }^{''}
+ \left( V_M + V_B + V_{L}^{*} \right)  \psi _{\nu } 
=\epsilon _{ \nu } \psi _{\nu }   \quad,
\label{eq:SL}
\end{equation}
where 
\begin{equation}
V_{L}^{*} (r) =  V_L(r)  \left [ 1 - \frac{1}{2} \left( 
\frac{\mu(\infty )  } {\mu(r) } 
\right)^{2}
\right] 
\label{eq:ULRtot}
\end{equation}
is attractive everywhere.

\begin{figure}[h!]
\begin{center}
\includegraphics[width=60mm]{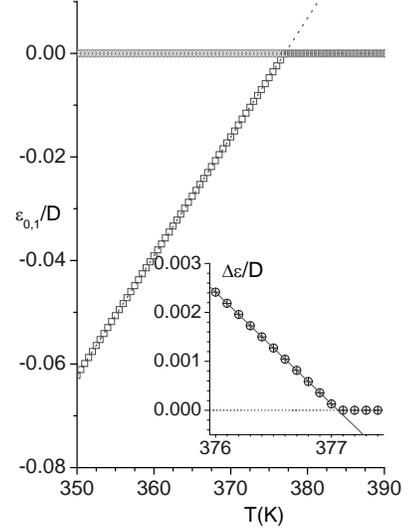} 
\caption{ The numerically determined
two lowest eigenvalues
of (\ref{eq:SL}), expressed in units of $D$. 
The onset shows details of the gap
in the region of the transition; no signs of
rounding are apparent at a sampling of
$\Delta T=0.1$~K. }
\label{fig:e0e1ODE}
\end{center}
\end{figure}

Eq. (\ref{eq:SL}) is a key result. It can be trivially cast in standard
Sturm-Liouville form with a density function proportional to $\mu ^2$; the $r$
dependence of $\mu$ is crucial for obtaining quantitatively sensible results; 
the simultaneous presence of three terms in the potential energy prevents us
from solving (\ref{eq:SL}) exactly. A numerical solution \cite{imsl}
for the parameter values of
Table (\ref{parameters})
reveals a behavior very similar to the full TI solution of 
Section \ref{sec:transferint}. According to Fig.   \ref{fig:e0e1ODE}, 
the two lowest eigenvalues exhibit an almost
perfect intersection at a temperature 
sampling $\Delta T=0.1$~K. 
In addition, the differential equation turns out to
be an excellent approximation to the original TI. Thus, 
the estimated $T_D=370$~K is only a few percent
higher than the value obtained within the TI; other critical thermodynamic 
quantities of interest demonstrate
comparable, or better agreement, e.g. the transition 
enthalpy $\Delta H=T_D \Delta S=0.129$~eV (cf. 0.133~eV from TI), or
the jump in the specific heat, $0.8 k_B$ (cf. $0.7k_B$ from TI).
The apparent first-order transition has its origin in the 
fact that the thermally generated barrier has a substantially 
longer range than the Morse potential.
The analysis of Section
\ref{sec:transferint}
suggests that crossover to a continuous transition 
eventually occurs; however, for the 
values of the parameters relevant to DNA denaturation, observing this
exponential 
crossover would require a temperature resolution of better than 
1~mK. 
This estimate can be made by studying crossover phenomena with exactly
solvable "toy models"\cite{TheoToy} of the denaturation transition 
of the linear PB variety,
where the zero stacking limit is known to yield a second-order "underlying"
transition. In the case of (\ref{eq:SL}), the presence of an attractive
inverse square interaction raises the possibility of more complex behavior,
i.e.  confinement at all temperatures or crossover to another transition at
much higher temperatures. This is discussed below.
\subsection{The $S=0$ case}
In the limit $S \to 0$, the decay of the kernel (\ref{ApprKern}) is governed 
by 
(\ref{eq:OmApproxG}). This reduces 
(\ref{eq:SL}) to
\begin{equation}
- k_{B}T \frac{\kappa ^2}{4} \psi _{\nu }^{''}
+ \left( V_M  + V_{L}^{*}   \right)  \psi _{\nu } 
=\epsilon _{ \nu } \psi _{\nu }   \quad,
\label{eq:SL0st}
\end{equation}
where 
\begin{equation}
V_{L}^{*} (r) =  -k_B T\frac{\kappa ^2}{16r^2}
\quad. 
\label{eq:U_0st}
\end{equation}
In the absence of stacking, the system is subject to the Morse potential
and the long-range attraction 
(\ref{eq:U_0st}); the point to note is that the 
attractive force is linearly dependent on the temperature, just as
the coefficient of the 2nd derivative in (\ref{eq:SL0st}); 
consequently, if $D=0$, the system will either have a bound 
state or not, according to the value of the coefficient in the
denominator of (\ref{eq:U_0st}). The value 16 is marginal; if
the interaction had been stronger, one would have confinement at
all temperatures; the Morse potential, being of short range, could
not change that; in other words, one would obtain a near-transition at
a temperature controlled by the Morse potential, but then the long-range
attraction would prevent dissociation at all temperatures. 
We have verified this by
numerically solving (\ref{eq:SL0st}). For weaker attractions
(value of the coefficient 16 or higher in these units),  numerical work
suggests that the transition becomes higher than second order; 
however,
numerical accuracy is not sufficient to determine the detailed behavior.
It is possible to guess what happens by substitutng the Morse
potential by a narrow well, i.e. the total potential in (\ref{eq:SL0st})
being equal to $-D$ for $R_0<r<R_0+1/a$ and equal (\ref{eq:U_0st})
for larger $r$; this case is exactly solvable and shows that
although the shift in the value of the critical point 
is less than 1 \% ,
the nature of the transition is  
radically transformed:
the vanishing of the lowest eigenvalue is now 
of the Kosterlitz-Thouless type
\begin{equation}
\epsilon _{0} \propto - e^{- const./(T_{D}-T) }
\quad.
\label{eq:KT}
\end{equation}

\begin{acknowledgments}
We thank Franco Bagnoli, Simona Cocco, Thierry Dauxois and Stefano
Ruffo  for useful discussions.  S.L. acknowledges partial financial
support from the Region Rh\^one-Alpes through the {\it Bourse
d'Accueil} nr. 00815559. Part of this work has been supported by the EU
Contract No. HPRN-CT-1999-00163 (LOCNET network).
\end{acknowledgments}


\begin{thebibliography}{99}
\bibitem{note} Following  Ref.~\cite{Coc99b} we also make the further
simplification to neglect the curvature terms originally introduced in
Ref.~\cite{Bar99a}. This also eases the determination of the model
parameters on the basis of the experimental data.
\bibitem{Symplectic}
For constant-energy simulation, it would be more proper to use a
symplectic algorithm. However is implementation would be awkward for
non-standard Lagrangians like ours as it would require implicit
schemes. Therefore we preferred to avoid this technical difficulty.
\bibitem{imsl}
We have used the IMSL subroutine DSLEIG in the semiinfinite
interval  $9.7$~\AA$ < r < \infty $.
\end{thebibliography}
\end{document}